\documentclass[preprint,aps,draft]{revtex4}
\usepackage{amssymb}

\begin{document}


\def\eqref#1{(\ref{#1})}
\def\eqrefs#1#2{(\ref{#1}) and~(\ref{#2})}
\def\eqsref#1#2{(\ref{#1}) to~(\ref{#2})}
\def\sysref#1#2{(\ref{#1})--(\ref{#2})}

\def\Eqref#1{Eq.~(\ref{#1})}
\def\Eqrefs#1#2{Eqs.~(\ref{#1}) and~(\ref{#2})}
\def\Eqsref#1#2{Eqs.~(\ref{#1}) to~(\ref{#2})}

\def\EQ #1\doneEQ{\begin{equation} #1 
\end{equation}}
\def\EQs #1\doneEQs{\begin{eqnarray} #1 
\end{eqnarray}}

\def\Ref#1{Ref.\cite{#1}}
\def\Refs#1{Refs.\cite{#1}}


\def\eqtext#1{\hbox{\rm{#1}}}

\def\mstrut{\mathstrut}
\def\hp#1{\hphantom{#1}}
\def\smallfrac#1#2{\textstyle{{#1}\over{#2}}}

\def\mixedindices#1#2{{\mstrut}^{\mstrut #1}_{\mstrut #2}}
\def\downindex#1{{\mstrut}^{\mstrut}_{\mstrut #1}}
\def\upindex#1{{\mstrut}_{\mstrut}^{\mstrut #1}}
\def\downupindices#1#2{{\mstrut}_{\mstrut #1}^{\hp{#1}\mstrut #2}}
\def\updownindices#1#2{{\mstrut}^{\mstrut #1}_{\hp{#1}\mstrut #2}}

\def\der#1{\partial\downindex{#1}}
\def\covder#1{\nabla\downindex{#1}}
\def\perpcovder#1{\nabla\mixedindices{\perp}{#1}}

\def\Sder#1{{\cal D}\downindex{#1}}
\def\covSder#1{\nabla\mixedindices{S}{#1}}
\def\perpSder#1{{\cal D}\mixedindices{\perp}{#1}}
\def\covDder#1{\covder{\Sigma}\downindex{#1}}

\def\Dder#1{D\downindex{#1}}
\def\slDder#1{
\setbox1=\hbox{$D$}
{\hbox to\wd1{$\box1\mkern-11mu\mathchar'57$\hfill}}_{#1} }

\def\slSder#1{
\setbox1=\hbox{$\cal D$}
\setbox2=\hbox{$\mathchar'57$}
{\hbox to\wd1{$\box1\mkern-11mu\raise1pt\box2$\hfill}}_{#1} }
\def\slcovder#1{
\setbox1=\hbox{$\nabla$}
\setbox2=\hbox{$\mathchar'57$}
{\hbox to\wd1{$\box1\mkern-12mu\raise1pt\box2$\hfill}}_{#1} }

\def\ethD#1{
\setbox1=\hbox{$\cal D$}
\setbox2=\hbox{$\mathchar'55$}
{\hbox to\wd1{$\box1 \mkern-12mu\raise2pt\box2$\hfill}}_{#1} }

\def\bslSder#1{
\setbox1=\hbox{$\cal D$}
\setbox2=\hbox{$\backslash$}
{\hbox to\wd1{$\box1\mkern-10mu\raise1pt\box2$\hfill}}_{#1} }
\def\bslethD#1{
\setbox1=\hbox{$\cal D$}
\setbox2=\hbox{$\mathchar'55$}
\setbox3=\hbox{$\backslash$}
{\hbox to\wd1
{$\box1 \mkern-12mu\raise2pt\box2\mkern-4mu\raise1pt\box3$\hfill}}_{#1} }

\def\g#1#2{g\downupindices{#1}{#2}}
\def\metric#1#2{\sigma\downupindices{#1}{#2}}
\def\perpmetric#1#2{\sigma\mixedindices{\perp}{#1}\upindex{#2}}
\def\vol#1#2{\epsilon\downupindices{#1}{#2}}
\def\perpvol#1#2{\epsilon\mixedindices{\perp}{#1}\upindex{#2}}
\def\coordvol#1#2{\varepsilon\downupindices{#1}{#2}}
\def\e#1#2{e_{#1}\upindex{#2}}
\def\inve#1#2{e^{#1}\downindex{#2}}
\def\vole#1#2{{\ast(e\wedge e)}\downupindices{#1}{#2}}

\def\h#1#2{h\downupindices{#1}{#2}}
\def\K#1#2{K\downupindices{#1}{#2}}
\def\trK{K}
\def\R#1#2{R\downupindices{#1}{#2}(\h{}{})}
\def\a#1#2{a\downupindices{#1}{#2}}

\def\x#1#2{x\mixedindices{#1}{#2}}
\def\id#1#2{\delta\mixedindices{#2}{#1}}

\def\curv#1#2{R\downupindices{#1}{#2}}
\def\scurv{R}
\def\ricci{Ric}
\def\conx#1#2{\Gamma\mixedindices{#2}{#1}}
\def\T#1#2{T\downupindices{#1}{#2}}

\def\curvS#1#2{{\cal R}\downupindices{#1}{#2}}
\def\scurvS{{\cal R}}
\def\gausscurvS{\chi}
\def\k#1#2{k\mixedindices{#2}{#1}}
\def\trk{\kappa}
\def\shear#1#2{{\cal K}\mixedindices{#2}{#1}}
\def\kperp#1#2{k_\perp\mixedindices{#2}{#1}}
\def\trkperp{\trk_\perp}
\def\norScurv#1#2{{\cal R}\mixedindices{\perp}{#1}\upindex{#2}}
\def\norscurvS{{\cal R}^\perp}

\def\perpcurv#1#2{R\mixedindices{\perp}{#1}\upindex{#2}}
\def\perpconx#1#2{{\cal J}\mixedindices{#1}{#2}}

\def\m{m}
\def\cm{\bar m}

\def\outnull{\ell}
\def\innull{n}

\def\frame#1#2{\vartheta\mixedindices{#2}{#1}}
\def\nframe#1#2{\vartheta\mixedindices{#2}{#1}}
\def\outnframe#1#2{\vartheta\mixedindices{+#1}{#2}}
\def\innframe#1#2{\vartheta\mixedindices{-#1}{#2}}
\def\volframe#1#2{{\ast(\vartheta\wedge\vartheta)}\downupindices{#1}{#2}}

\def\indS#1{{\underline #1}}

\def\mcurv{H}
\def\perpmcurv{H_\perp}
\def\twist{\varpi}
\def\perptwist{\varpi^\perp}
\def\P{P}
\def\absmcurv{|\mcurv|}
\def\absperpmcurv{|\perpmcurv|}
\def\unitmcurv{\hat H}
\def\unitperpmcurv{{\hat H}_\perp}

\def\Pvec#1#2{\P\mixedindices{#1}{#2}}
\def\Hvec#1#2{\mcurv\mixedindices{#1}{#2}}
\def\perpHvec#1#2{\mcurv\mixedindices{#1}{\perp #2}}
\def\twistvec#1#2{\twist\mixedindices{#1}{#2}}

\def\nullmcurv#1{\mcurv_{#1}}

\def\TS{T(S)}
\def\TperpS{T(S)^\perp}
\def\coTS{T^*(S)}
\def\coTperpS{T^*(S)^\perp}
\def\TM{T(M)}
\def\coTM{T^*(M)}
\def\ThM{T(\Sigma)}
\def\coThM{T^*(\Sigma)}

\def\hook{\rfloor}
\def\Lie#1{\pounds_{#1}}
\def\unitop{\openone}
\def\proj#1#2{{\cal P}^{#1}_{#2}}

\def\tr{{\rm tr}}
\def\ad{\dagger}

\def\i{{\rm i}}

\def\d{{\rm d}}

\def\diag{{\rm diag}}

\def\flow{\xi}
\def\flowvec#1#2{\xi\mixedindices{#1}{#2}}
\def\flux{{\cal F}}

\def\gmatr#1#2{\gamma^{#2}_{#1}}
\def\hyperpsi{\psi\downindex{\Sigma}}
\def\spin#1#2{\psi\mixedindices{#2}{#1}}
\def\varspin#1#2{\varphi\mixedindices{#2}{#1}}
\def\cspin#1#2{\bar\psi\mixedindices{#2}{#1}}
\def\cvarspin#1#2{\bar\varphi\mixedindices{#2}{#1}}
\def\ospin#1#2{o\mixedindices{#2}{#1}}
\def\ispin#1#2{\iota\mixedindices{#2}{#1}}
\def\ocspin#1#2{\bar o\mixedindices{#2}{#1}}
\def\icspin#1#2{\bar\iota\mixedindices{#2}{#1}}

\def\flt{{\rm flat}}
\def\twt{{\rm twist}}
\def\Mink{{\rm Mink}}
\def\hor{{\rm hor}}

\def\const{{\rm const}}
\def\Rnum{{\mathbb R}}
\def\cc{{\rm c.c.}}

\def\ie/{i.e.}
\def\etc/{etc.}

\title{ Spinor derivation of quasilocal mean curvature mass\\
in General Relativity }

\author{Stephen C. Anco}
\email{sanco@brocku.ca}
\affiliation{%
Department of Mathematics, Brock University, St Catharines,
Ontario, L2S 3A1, Canada
}

\date{\today}

\begin{abstract}
A spinor derivation is presented for quasilocal mean-curvature mass
of spacelike 2-surfaces in General Relativity. 
The derivation is based on the Sen-Witten spinor identity 
and involves the introduction of novel nonlinear boundary conditions
related to the Dirac current of the spinor at the 2-surface
and the tangential flux of a boundary Dirac operator,
as well the use of a spin basis adapted to the mean curvature frame of
the 2-surface normal space. 
This setting may provide an alternative approach to a 
positivity proof for mean-curvature mass based on showing that
Witten's equation admits a spinor solution satisfying the proposed
nonlinear boundary conditions. 
\end{abstract}

\maketitle

\section{ Introduction }

In General Relativity 
there is substantial interest in extending Witten's spinor proof
for positivity of the ADM mass \cite{Witten,Choquet-Bruhat}
in asymptotically flat spacetimes 
to the setting of a quasilocal mass for spacelike 2-surfaces. 
Two major difficulties with such a proof have been, 
firstly, how to relate the 2-surface side of the Sen-Witten spinor identity
to a well-defined quasilocal mass or energy expression  
with satisfactory properties \cite{criteria};
and secondly, what boundary conditions to impose in Witten's equation 
on the spinor at the 2-surface. 
Resolving these obstacles could provide a new proof of positivity
for some of the several known quasilocal mass-energy definitions
(see \cite{survey} for a comprehensive review)
or perhaps even suggest a more fully satisfactory definition
based directly on spinors. 

In this paper, 
a formal spinor derivation is presented for quasilocal mean-curvature mass 
and its variants studied in recent work \cite{Kijowski,Lau,Epp,LiuYau,Anco}.
The key ideas here will involve 
the introduction of nonlinear boundary conditions for Witten's equation
that are related to the Dirac current of the spinor at the 2-surface 
and also to the tangential flux of a Dirac operator on the 2-surface, 
combined with a spin basis adapted to
the mean curvature frame of the 2-surface normal bundle,
which will be used for evaluating the Sen-Witten identity
in section~II. 
Remarks toward a positivity argument are made in section~III, 
by showing that the positivity of the mean-curvature mass 
reduces to existence of a spinor solution of Witten's equation 
subject to the proposed nonlinear boundary conditions
on a spacelike hypersurface that spans the 2-surface. 
This argument extends to a Hamiltonian form of mean-curvature mass
discussed in section~IV. 
In addition, 
the standard treatment of horizons as inner boundaries 
for Witten's equation using chiral boundary conditions \cite{GHHP}
(at the horizon) 
is shown to carry through in the present setting for mean-curvature mass
in section~V. 

By way of concluding remarks in section~VI, 
a link is pointed out between mean-curvature mass
and a positive quasilocal spinorial mass
defined by introducing 
a mean curvature variant of chiral boundary conditions 
(in place of the previous nonlinear boundary conditions) 
for which the existence and uniqueness of solutions of Witten's equation
can be rigorously established \cite{BartnikChrusciel}.

A 4-spinor formalism \cite{GHHP}
will be used throughout, 
employing Dirac spinors $\psi$ and orthonormal frames $\e{a}{}$
and coframes $\e{}{a}$, $a=0,1,2,3$.
Key equations will be summarized at the end in 2-spinor index form. 
Numerical gamma matrices $\gmatr{a}{}$ attached to an orthonormal frame 
will satisfy the Clifford algebra
$\gmatr{a}{}\gmatr{b}{} + \gmatr{b}{}\gmatr{a}{} = 2\g{ab}{}\unitop$
in terms of frame components $\g{ab}{} =\diag(-1,+1,+1,+1)$
of the spacetime metric tensor
$g=\g{ab}{} \e{}{a}\otimes\e{}{b}$,
such that $\gmatr{0}{}$ is anti-hermitian 
while $\gmatr{i}{}$ is hermitian. 
With these standard conventions
the Dirac current $\bar\psi \gamma(e)\psi \equiv -\flow(\psi)$
of a spinor is a timelike past-pointing vector field
such that $\flowvec{0}{}(\psi)=|\psi|^2$ is the norm of $\psi$,
where $\gamma(e) = \gmatr{}{a}\e{a}{}$ 
is the dual to the soldering form. 
The covariant derivative operator $\covder{}$ acting on spinor fields
will be given by 
$\covder{} = \der{} +\frac{1}{4} \gmatr{ab}{} \conx{}{ab}$
with $\gmatr{ab}{} \equiv \frac{1}{2} [\gmatr{a}{},\gmatr{b}{}]$
where $\der{}$ denotes the coordinate derivative operator
and $\conx{}{ab}$ denotes the spin connection 1-form, 
given in terms of the coframe by 
$\conx{a}{\hp{a}b} = \covder{a}\e{}{b} -\e{a}{}\hook(\d\e{}{b})$.
For a spacelike 2-surface $S$ 
spanned by a smooth spacelike hypersurface $\Sigma$ 
in a spacetime $(M,g)$,
the orthonormal frame $\e{a}{}$ will be adapted to $\Sigma$ and $S$
so that  
$\e{0}{}$ is normal to $\Sigma$
and future-pointing, 
$\e{1}{}|_S$ is orthogonal to $S$ in $\Sigma$
and outward directed, 
$\e{\indS{a}}{}|_S$ is tangential to $S$ for $\indS{a}=2,3$.
The 2-surface metric tensor will be denoted 
$\sigma = g|_S 
= \g{\indS{a}\indS{b}}{} (\e{}{\indS{a}}\otimes\e{}{\indS{b}})|_S$. 
Spacetime coordinates will also be adapted to $\Sigma$ and $S$
by use of a coordinate derivative $\der{\Sigma}$ on $\Sigma$
and its restriction $\der{S}$ on $S$. 
(In general subscripts $S,\Sigma$ will denote a restriction to 
tangent spaces $T(S),T(\Sigma)$ or 
cotangent spaces $T^*(S),T^*(\Sigma)$, as appropriate.)

\section{ Mean-curvature mass }

To proceed, 
introduce
\EQ
\mcurv \equiv \trk(\e{1}{}) \e{1}{} - \trk(\e{0}{}) \e{0}{}
\doneEQ
which defines 
the mean curvature vector \cite{ONeillI,Chen,Anco} of $S$ in $M$,
where $\trk(\e{0}{}),\trk(\e{1}{})$ are the extrinsic scalar curvatures 
(trace of the second fundamental form) of $S$
relative to the normal frame $\{\e{0}{},\e{1}{}\}$ of $S$,
and write \cite{Anco}
\EQ
\perpmcurv \equiv \trk(\e{1}{}) \e{0}{} - \trk(\e{0}{}) \e{1}{} =*\mcurv
\doneEQ
for the dual of $\mcurv$ in the normal space of $S$. 
The absolute norm
\EQ 
\absmcurv =\absperpmcurv \equiv
\sqrt{| \trk(\e{1}{})^2 - \trk(\e{0}{})^2 |}
\doneEQ
defines the scalar mean curvature of $S$ in $M$.
Note the vectors $\mcurv,\perpmcurv$, as well as the scalar $\absmcurv$,
are independent of choice of a normal frame 
(they display invariance under boosts and reflections of $\e{0}{},\e{1}{}$)
and hence are geometrically well-defined given just the 2-surface 
and its extrinsic geometry in spacetime.  
These vectors satisfy the convention that if $M$ is Minkowski space
and $S$ is a convex 2-surface lying in a spacelike hyperplane
then $\mcurv$ is spacelike and outward directed,
while $\perpmcurv$ is timelike and future-pointing,
and the norm of $\mcurv$ 
agrees with the Euclidean mean curvature of $S$.
An important geometric property of $\perpmcurv$ is that $S$ 
has no expansion in this direction in spacetime
\EQ\label{liedermcurv}
\trk(\perpmcurv) = \frac{1}{2} \tr( \Lie{\perpmcurv} \sigma )=0 ,
\doneEQ
namely the extrinsic scalar curvature of $S$ in the $\perpmcurv$ direction 
vanishes,
while the extrinsic scalar curvature of $S$ in the direction 
orthogonal to $\perpmcurv$ 
is given by the norm of $\mcurv$,
\EQ
\trk(\mcurv) = \frac{1}{2} \tr( \Lie{\mcurv} \sigma )= \mcurv\cdot\mcurv .
\doneEQ
Hereafter $S$ is assumed to have $\mcurv$ spacelike,
so $\perpmcurv$ is timelike and 
$\absmcurv =\sqrt{\mcurv\cdot\mcurv}
= \absperpmcurv=\sqrt{-\perpmcurv\cdot\perpmcurv}$ is non-negative. 

The quasilocal mean-curvature mass of $S$
is defined by the surface integral 
\EQ\label{mcurvmass}
E(S;\sigma) \equiv \frac{1}{8\pi}
\int_S ( \absmcurv_\flt -\absmcurv ) \d S
\doneEQ
where $\absmcurv_\flt$ is the Euclidean mean curvature given by 
an isometric embedding of $(S,\sigma)$ 
into a spacelike hyperplane in Minkowski space.
(By Weyl's theorem \cite{weylembed}, such an embedding exists 
and is unique up to rigid motions if $S$ has positive Gaussian curvature.)
As shown in recent work \cite{Anco},
this mass has good geometric properties:
it agrees with the ADM mass in a large sphere limit at spatial infinity
in asymptotically flat spacetimes,
and it is bounded below by (twice) the irreducible mass 
$M_{\rm irr} \equiv\sqrt{A/16\pi}$ 
at apparent horizons 
(namely, 2-surfaces whose scalar mean curvature vanishes, 
$\absmcurv=0$ on $S$)
where $A$ denotes the area of $S$. 
Most importantly, the mean-curvature mass is non-negative 
in all spacetimes that satisfy the dominant energy condition, 
as proved in \Ref{LiuYau}.
This positivity indicates that mean-curvature mass may be very useful
in the context of geometric analysis problems in General Relativity. 
On the other hand, the physical meaning of this mass is not so clear
because it fails to vanish \cite{survey} 
for certain 2-surfaces (not lying in hyperplanes) in Minkowski space. 

The starting point for a spinor derivation is the Witten-Nester 2-form 
\cite{Witten,Nester,Choquet-Bruhat}
given by 
$\bar\psi \gmatr{5}{} \gamma(e) \wedge \covder{}\psi$,
where $\gamma(e)= \e{}{a}\gmatr{a}{}$ is the soldering form,
and $\gmatr{5}{} \equiv \gmatr{0}{}\gmatr{1}{}\gmatr{2}{}\gmatr{3}{}$. 
Write
\EQ
\covDder{} = 
\der{\Sigma}
+\frac{1}{4} \gmatr{ab}{} \conx{\Sigma}{ab} ,\quad
\Sder{} =
\der{S}+\frac{1}{4} \gmatr{\indS{a}\indS{b}}{} \conx{S}{\indS{a}\indS{b}} 
\doneEQ
respectively for the spatial covariant derivative
and the 2-surface intrinsic covariant derivative
acting on spinors; 
a slash will be used to denote the contraction of these derivative operators
with the soldering form $\gamma(e)$. 
Likewise
\EQ
\slcovder{S} \equiv \gmatr{}{\indS{a}} \covder{\indS{a}} =
\slSder{} +\frac{1}{2}\trk(\e{0}{})\gmatr{}{0}
+\frac{1}{2}\trk(\e{1}{})\gmatr{}{1}
+\twistvec{}{\indS{a}}(e) \gmatr{}{\indS{a}} \gmatr{}{1}\gmatr{}{0}
\doneEQ
denotes a Dirac operator associated to $S$ as a 2-surface sitting in $M$. 
Recall, 
$\trk(\e{0}{}) \equiv \e{}{\indS{a}}\cdot \covder{\indS{a}} \e{0}{}$
and 
$\trk(\e{1}{}) \equiv \e{}{\indS{a}}\cdot \covder{\indS{a}} \e{1}{}$
are the extrinsic scalar curvatures of $S$ in the normal directions 
$\{\e{0}{},\e{1}{}\}$,
and $\twistvec{}{\indS{a}}(e) \equiv \e{1}{}\cdot\covder{\indS{a}}\e{0}{}$
is the twist of the normal frame in the tangential directions
$\{\e{\indS{a}}{}\}$. 
Then, for any smooth spinor $\psi$
and smooth hypersurface $\Sigma$ with a 2-surface boundary $S=\partial\Sigma$, 
integration of the Witten-Nester 2-form over $S$
followed by use of Stokes' theorem, 
along with some gamma matrix algebra, 
gives the well-known Sen-Witten identity:
\EQ
\int_{S} ( \psi^\ad \gmatr{}{1}\slcovder{S}\psi +\cc ) \d S
= 2\int_\Sigma ( 
-\covDder{}\psi^\ad \cdot \covDder{}\psi
+ \T{0a}{} \psi^\ad\gmatr{}{0}\gmatr{}{a} \psi
+ (\slcovder{\Sigma}\psi)^\ad (\slcovder{\Sigma} \psi) 
) \d\Sigma
\doneEQ
($\cc$ stands for the complex conjugate of the preceding term)
where 
\EQ
8\pi\T{ab}{} \equiv \ricci(\e{a}{},\e{b}{}) -\frac{1}{2} \g{ab}{}\scurv
\doneEQ
defines the stress-energy of $(M,g)$ in terms of 
the spacetime Ricci curvature tensor,
with $\scurv=\g{}{ab} \ricci(\e{a}{},\e{b}{})$
being the scalar curvature. 

For the sequel 
it will be natural to introduce the following geometric spinor operators
(see \Ref{survey,BartnikChrusciel} for a summary of relevant 
mathematical background):
\EQ
\ethD{\indS{a}} \equiv 
\Sder{\indS{a}} +\frac{1}{2} \gmatr{}{1}\gmatr{}{0}\twistvec{}{\indS{a}}(e) 
\doneEQ
is a linear combination of the spinorial edth operators \cite{edth,GHHP};
and 
\EQ
\bslSder{} \equiv 
\gmatr{}{1} \gmatr{}{\indS{a}} \Sder{\indS{a}} ,\quad
\bslethD{} \equiv 
\gmatr{}{1} \gmatr{}{\indS{a}} \ethD{\indS{a}}
\doneEQ
are boundary flux operators 
related to the Dirac operator in Witten's equation \cite{GHHP,Herzlich}
as seen later. 
(Note that 
$\gmatr{}{1}\gmatr{}{\indS{a}}$ 
gives an alternate representation of the Clifford algebra over $\TS$,
the more obvious representation being $\gmatr{}{\indS{a}}$
induced from the Clifford algebra over $T(M)$).
These operators $\bslethD{}$ and $\bslSder{}$ 
are frame-invariant and depend just on the intrinsic geometry of $S$. 

Now let $\Sigma_\perp$ be a smooth spacelike hypersurface spanning $S$
such that it is 
orthogonal to the dual mean curvature vector $\perpmcurv$ at $S$ in $M$.
The corresponding adapted orthonormal frame $\e{a}{}$ has 
$\e{0}{}|_S=\unitperpmcurv \equiv \absmcurv^{-1}\perpmcurv$ 
normal to $\Sigma_\perp$ at $S$, 
$\e{1}{}|_S=\unitmcurv \equiv \absmcurv^{-1}\mcurv$  
orthogonal to $S$ in $\Sigma_\perp$,
$\e{\indS{a}}{}|_S$ tangential to $S$ as before;
this frame will be distinguished by placing a hat atop 
frame-dependent spinor operators, vectors and 1-forms. 
Let 
\EQ
\twist \equiv \hat\twist(e) = 
(\unitmcurv \cdot \covder{\indS{a}} \unitperpmcurv) 
\e{}{\indS{a}}
\doneEQ
which denotes the twist covector of the mean curvature frame of $S$. 
In this geometrically preferred frame 
the covariant edth operator and boundary flux operator are given by 
\EQ
\widehat{\ethD{}} 
= \Sder{} +\frac{1}{2} \gmatr{}{1}\gmatr{}{0} \twist ,\quad
\widehat{\bslethD{}} 
= \gmatr{}{1}\gmatr{}{\indS{a}} \widehat{\ethD{\indS{a}}} ,
\doneEQ
while the boundary Dirac operator takes the form
\EQ
\slcovder{S} = \gmatr{}{1}(\widehat{\bslethD{}} + \frac{1}{2}\absmcurv)
\doneEQ
due to the property \eqref{liedermcurv} of the mean curvature frame. 
In particular, 
$\widehat{\ethD{}},\widehat{\bslethD{}}$, 
and $\twist,\mcurv,\perpmcurv$
all are geometrically well-defined given just the 2-surface $S$
and its extrinsic geometry in spacetime.  
The Sen-Witten identity now becomes
\EQ\label{senwittenid}
\int_S
( \psi^\ad \widehat{\bslethD{}}\psi +\cc +\absmcurv |\psi|^2 ) \d S
= -2\int_{\Sigma_\perp} ( 
|\covDder{}\psi|^2 -\flow(\psi)\cdot \T{0}{}
- |\slcovder{\Sigma}\psi|^2 
) \d\Sigma
\doneEQ
with 
$|\covDder{}\psi| \equiv \sqrt{ \covDder{}\psi^\ad \cdot \covDder{}\psi }$
and
\EQ 
\T{0}{} \equiv 
\T{0}{a}\e{a}{} ,\quad
\flow(\psi) \equiv 
-\bar\psi \gmatr{}{a}\e{a}{}\psi , 
\doneEQ
as defined in the adapted orthonormal frame. 

The aim is now to relate the surface integral terms 
in the spinor identity \eqref{senwittenid} 
to a quasilocal mass expression. 
This will be accomplished through considering
the tangential Dirac current 
\EQ
\flow_\parallel(\psi_S) \equiv 
-(\bar\psi \gmatr{}{\indS{a}}\e{\indS{a}}{}\psi)|_S
\doneEQ
and the tangential flux 
\EQ
\flux(\psi_S) \equiv (\psi^\ad\bslSder{}\psi)|_S +\cc
\doneEQ
of $\psi_S \equiv \psi|_S$ 
connected with an isometric embedding of $S$ 
into a spacelike hyperplane in Minkowski space. 
Note $(\covder{}\perpmcurv)_\flt=0$
and hence $(\twist)_\flt=0$
holds on the embedded 2-surface $S_\flt$,
so thus $\ethD{\flt}=\Sder{}$
and 
$(\slcovder{S})_{\Mink} = \slSder{} +\frac{1}{2}\absmcurv_{\flt}\gmatr{}{1}$
using an orthonormal frame $(\e{a}{})_{\Mink}$
adapted to the hyperplane and the embedded 2-surface
(whose mean curvature frame $\{\unitperpmcurv,\unitmcurv\}_{\flt}$
then coincides with the adapted normal frame for $a=0,1$
and whose tangential frame pulls back to the one on $S$ for $a=2,3$).
With respect to this embedding, choose $\psi$ to be a parallel
(covariantly constant) spinor in Minkowski space,
$(\covder{}\psi)_{\Mink} =0$,
with
\EQ
\flowvec{a}{}(\psi)=(|\psi|^2,0,0,0) ,\quad
|\psi|= \const . 
\doneEQ
More geometrically, the embedded Dirac current vector of $\psi$ at $S_{\flt}$
is aligned with the dual mean curvature vector of $S_{\flt}$,
\EQ
(\flow(\psi_S)\wedge\perpmcurv)_{\flt}=0 , 
\doneEQ
and has constant absolute norm,
\EQ\label{normfactor} 
|\flow(\psi_S)| = |\psi|^2= \const .
\doneEQ
Moreover, $\psi$ satisfies the boundary Witten equation 
\EQ\label{witteneqS}
(\slcovder{S} \psi_S)_{\Mink} = 
\gmatr{}{1}( \bslSder{} \psi_S 
+\frac{1}{2} \absmcurv_{\flt} \psi_S )
=0
\doneEQ
on the embedded 2-surface $S_{\flt}$, 
and so the tangential flux
\EQ\label{flux}
\flux(\psi_S) = 
- \absmcurv_{\flt} |\psi_S|^2
\doneEQ
is a constant multiple of the Euclidean mean curvature of $S_{\flt}$. 
Note, for comparison, 
the boundary Witten equation on $S$ itself in the spacetime $M$ 
would look like
\EQ
\slcovder{S} \phi 
= 
\slSder{}\phi +\frac{1}{2} \trk(\e{1}{})\gmatr{}{1} \phi
+\frac{1}{2}\trk(\e{0}{})\gmatr{}{0}\phi 
+\twistvec{}{\indS{a}}(e)\gmatr{}{\indS{a}}\gmatr{}{1}\gmatr{}{0}\phi
= 
\gmatr{}{1}( \bslethD{}\phi +\frac{1}{2} \trk(\e{1}{})\phi
+\frac{1}{2}\trk(\e{0}{})\gmatr{}{1}\gmatr{}{0}\phi )
=0 
\doneEQ
relative to a general orthonormal frame $\e{a}{}$,
for any spinor $\phi$. 
Now, substitution of equations \eqrefs{normfactor}{flux}
into the Sen-Witten identity \eqref{senwittenid}
yields a main result:
\EQs
\int_S \psi^\ad\widehat{\bslethD{}}\psi +\cc +\absmcurv |\psi|^2 \d S
&& = \int_S ( 
\flux(\psi_S) 
+\twist\cdot \flow_\parallel(\psi_S)
+\absmcurv |\flow(\psi_S)|
)\d S 
\nonumber\\
&& = -|\psi_S|^2 \int_S ( \absmcurv_\flt -\absmcurv ) \d S ,\quad
|\psi_S| =\const . 
\doneEQs
Note there is no loss of generality in scaling $\psi$ 
by a constant factor so that $|\psi_S| =1$.

Theorem~1:
Let $S$ be a spacelike 2-surface whose mean curvature vector $\mcurv$ 
is spacelike and whose Gaussian curvature is positive. 
Fix a Dirac spinor $\psi$ in Minkowski space satisfying
boundary conditions 
$\flux(\psi_S) = -\absmcurv_{\flt}$, 
$\flow_\parallel(\psi_S)=0$, and $|\psi_S|=1$ 
on $S$ embedded into a spacelike hyperplane. 
(In particular such boundary conditions hold when $\psi$ is any
parallel spinor, $(\covder{}\psi)_{\flt}=0$,
aligned and normalized so that 
$\flow(\psi_S)_{\flt}=(\unitperpmcurv)_{\flt}$
holds in the embedding.)
Then the surface integral terms in the Sen-Witten identity 
are a multiple of the mean-curvature mass \eqref{mcurvmass} of $S$:
\EQ\label{maineq}
8\pi E(S;\sigma) 
= -\int_S \psi^\ad\widehat{\bslethD{}}\psi +\cc +\absmcurv |\psi|^2 \d S . 
\doneEQ

\section{ Remarks on positivity }

This spinor derivation of the mean curvature mass 
allows the possibility of modifying Witten's positivity argument 
as follows. 
Impose on $\psi$ the Witten equation
\EQ\label{witteneq}
\slcovder{\Sigma}\psi =0
\doneEQ
subject to the nonlinear boundary conditions
\EQs
&&
\flux(\psi_S) = \flux(\phi) = - \absmcurv_{\flt} ,
\label{meancurvbc3}\\
&& 
\flow_\parallel(\psi_S) = \flow_\parallel(\phi) = 0 ,
\label{meancurvbc1}\\
&&
|\psi_S| = |\phi| = 1 ,
\label{meancurvbc2}
\doneEQs
under the isometric embedding of $(S,\sigma)$
into a spacelike hyperplane in Minkowski space, 
where $\phi$ is the restriction to $S$ of a parallel spinor 
whose Dirac current vector is aligned with 
the dual mean curvature vector of $S$ in the embedding, 
$\flow(\phi)_{\flt}= (\unitperpmcurv)_{\flt}$. 
Note $\phi$ then satisfies the boundary Witten equation (in Minkowski space)
\EQ\label{DiracSeq}
(\slcovder{S}\phi)_{\Mink}
= \gmatr{}{1}( \bslSder{}\phi +\frac{1}{2}\absmcurv_{\flt}\phi ) =0
\doneEQ
such that the boundary Dirac current vector has unit absolute norm
\EQ\label{DiracSnorm} 
|\flow(\phi)| = |\phi|^2= 1 . 
\doneEQ

These boundary conditions specify 
the tangential Dirac current of $\psi$ on $S$
and the tangential flux of $\psi$ on $S$,
along with the norm of $\psi$ 
(which is given by 
the projection of the Dirac current vector of $\psi$
in the timelike mean curvature direction $\perpmcurv$ at $S$,
$\flow(\psi_S)\cdot\unitperpmcurv=-|\psi_S|^2$). 
The boundary value problem 
given by equations \sysref{witteneq}{meancurvbc2} on $\psi$
constitutes a first-order elliptic PDE system 
with the right number of boundary conditions 
(half of the degrees of freedom of the spinor). 
To give a more rigorous indication of well-posedness, 
it would be natural to first study the linearized equations
as obtained in a weak gravitational limit
(where the spacetime metric is a perturbation of the Minkowski metric). 
This would yield a linear, non-homogeneous boundary value problem 
in Minkowski space to which standard Green's function techniques 
could be applied to verify if there exist solutions $\psi$. 
Such an analysis will be left for investigation elsewhere,
and the rest of the argument will now be formal. 

From Witten's equation \eqref{witteneq} 
in the nonlinear boundary value problem, 
since the hypersurface $\Sigma_\perp$ is orthogonal to 
the timelike mean curvature normal of $S$, 
the normal covariant derivative of $\psi$ at $S$ in $\Sigma_\perp$ 
is given by
\EQ
(\covder{1}\psi)|_S 
= -\gmatr{}{1}\slcovder{S}\psi_S
= -(\widehat{\bslethD{}}\psi_S +\frac{1}{2}\absmcurv\psi_S)
\doneEQ
and consequently the normal flux of $\psi$ on $S$ is precisely the same
as the Sen-Witten surface integral density expression.
In particular, 
under the boundary conditions \sysref{meancurvbc3}{meancurvbc2},
\EQ
(\psi^\ad\covder{1}\psi)|_S +\cc 
= -( \psi_S^\ad\widehat{\bslethD{}}\psi_S +\cc +\absmcurv |\psi_S|^2 )
= \absmcurv_{\flt} -\absmcurv
\doneEQ
so thus the normal flux of $\psi$ on $S$ is equal to 
the difference of the scalar mean curvature of $S$
(as a 2-surface in $M$)
and the Euclidean mean curvature of $S$
(as an embedded 2-surface in Minkowski space). 

As a result the spinor derivation of mean-curvature mass \eqref{maineq}
holds (similarly to theorem~1)
and is directly related to the normal flux of the spinor in Witten's equation
at the 2-surface in spacetime. 
Then equations \eqref{senwittenid} and \eqref{witteneq}
yield
\EQ
8\pi E(S;\sigma)
= 2 \int_{\Sigma_\perp} ( 
|\covDder{}\psi|^2 -\flow(\psi)\cdot \T{0}{} ) \d\Sigma
\geq 0
\label{mainineq}
\doneEQ
since
\EQ
-\flow(\psi)\cdot \T{0}{} 
\geq 0
\doneEQ
if the local energy-momentum vector field
$\T{0}{}$ associated with $\Sigma_\perp$ 
is timelike and future-pointing
(which is just the dominant energy condition \cite{HawkingEllis,Wald-book}
on the spacetime stress-energy tensor $\T{ab}{}$). 

Proposition~2:
Let $S$ be a 2-surface as in theorem~1. 
If a smooth solution $\psi$ of 
the Witten equation \eqref{witteneq} exists 
satisfying the nonlinear boundary conditions 
\sysref{meancurvbc3}{meancurvbc2} 
on a hypersurface $\Sigma_\perp$ spanning $S$
then the mean-curvature mass \eqref{mcurvmass} of $S$ is non-negative
provided the spacetime $(M,g)$ obeys the dominant energy condition.

\section{ Twist-free 2-surfaces and mean-curvature mass }

A 2-surface $S$ is said to be {\it convex} if 
its mean curvature vector $\mcurv$ is spacelike, 
or equivalently the dual mean curvature vector $\perpmcurv$ is timelike;
and said to be {\it twist-free} if the twist of these normal vectors 
$\twistvec{}{\indS{a}}(\mcurv) = \mcurv\cdot\covder{\indS{a}}\perpmcurv$ 
vanishes, 
so therefore a convex twist-free $S$ 
possesses a mean curvature normal frame 
$\{\unitperpmcurv,\unitmcurv\}$
whose twist is zero, $\twist=0$. 

For any such 2-surface, 
an elegant version of the spinor derivation 
can be formulated using 
$SU(2)$ spinors on a maximal spacelike hypersurface $\Sigma$.
Firstly, recall that the Witten equation on a maximal hypersurface
reduces to the 3-dimensional Witten equation 
by means of the spatial Dirac operator decomposition 
\EQ
\slcovder{\Sigma} 
= \slDder{} +\frac{1}{2} \gmatr{}{0} \tr(\K{}{}(\e{0}{}))
\doneEQ
with 
$\Dder{} = \der{\Sigma} +\frac{1}{4} \gmatr{ij}{} \conx{\Sigma}{ij}$
being the hypersurface covariant derivative,
where 
$h=g|_S = \g{ij}{}(\e{}{i}\otimes\e{}{j})|_\Sigma$
is the spatial metric tensor in terms of the spatial coframe $\e{}{i}$,
$i=1,2,3$, intrinsic to $(\Sigma,\h{}{})$,
and where $\K{}{}(\e{0}{})=\frac{1}{2}\Lie{\e{0}{}}\h{}{}$
is the second fundamental form of $\Sigma$ 
in terms of the spatial metric $\h{}{}$.
Here $\slDder{}$ is the spatial (3-dimensional) Dirac operator
associated to $\Sigma$. 
Secondly, recall that $SU(2)$ spinors, $\hyperpsi$, 
on a spacelike hypersurface are defined via projection operators 
$\proj{0}{\pm} = \frac{1}{2}( \unitop \pm \i \gmatr{}{0} )$
by the conditions 
$\proj{0}{+}\psi = \psi$ or equivalently $\proj{0}{-}\psi = 0$ 
applied to a Dirac spinor $\psi$
\cite{Sen,Wald-book}. 
This projection reduces the number of 
linearly independent real components of $\hyperpsi$ to four. 
A key property of these spinors is their compatibility with 
the spatial Dirac operator, 
\EQ
\slDder{} \proj{0}{\pm} = \proj{0}{\mp} \slDder{} , 
\doneEQ
as seen by using 
$[\proj{0}{\pm},\gmatr{}{0}]=0$
and 
$=\proj{0}{\pm}\gmatr{}{i} - \gmatr{}{i}\proj{0}{\mp} =0$. 

To set up the derivation of the mean-curvature mass,
assume $S$ is spanned by 
a maximal spacelike hypersurface $\Sigma_\perp$
orthogonal to $\perpmcurv$ at $S$. 
Suppose there exists an $SU(2)$ spinor $\hyperpsi$ 
satisfying the spatial Witten equation
\EQ\label{spatialdiraceq}
\slDder{} \hyperpsi =0
\doneEQ
and nonlinear boundary conditions \eqrefs{meancurvbc3}{meancurvbc2}
under an isometric embedding of $(S,\sigma)$
into a spacelike hyperplane in Minkowski space, 
where $\psi_S$ is the restriction of $\hyperpsi$ to $S$. 
Here $\phi$ is a parallel $SU(2)$ spinor in Minkowski space,
$(\covder{}\phi)_{\Mink}=0$,
with its Dirac current vector aligned with 
the dual mean curvature vector of $S$ in the embedding, 
$\flow(\phi)_{\flt}= (\unitperpmcurv)_{\flt}$. 
As before, 
$\phi$ thus satisfies the embedded boundary Witten equation \eqref{DiracSeq}
with the boundary Dirac current obeying the normalization \eqref{DiracSnorm}. 

The elegance of $SU(2)$ spinors now comes into play in linking 
the Sen-Witten identity to the mean-curvature mass in a Hamiltonian form
\cite{Anco} as follows. 
First, using the projection operators it can be easily shown that
the Dirac current of $\hyperpsi$ is orthogonal to $\Sigma_\perp$, 
\EQ\label{su2current}
\flow(\hyperpsi) = |\hyperpsi|^2 \e{0}{} .
\doneEQ
This identity gives a simple spinor parametrization of
the timelike mean curvature vector of $S$
since $\e{0}{}|_S =\unitperpmcurv$.

Theorem~3:
On a convex twist-free spacelike 2-surface $S=\partial\Sigma_\perp$,
the surface integral terms in the Sen-Witten identity on $\hyperpsi$
subject to boundary conditions \eqrefs{meancurvbc3}{meancurvbc2}
in the adapted mean curvature frame
are given by 
\EQ\label{senwittensu2}
-\int_S
( \hyperpsi{}^\ad \widehat{\slSder{}}\hyperpsi +\cc 
+\absmcurv |\hyperpsi|^2 ) \d S
= \int_S ( \absmcurv_\flt -\absmcurv ) \d S
= \int_S \flow\cdot\P \d S 
- \int_S (\flow\cdot\P)_{\flt} \d S 
\doneEQ
where $\P\equiv \perpmcurv +\twist$
is the symplectic vector \cite{paperI,paperII,errata} 
in a Hamiltonian formulation of the Einstein gravitational field equations
using a geometric time flow vector field \cite{Anco}
which is given at $S$ by the timelike mean curvature vector
$\flow \equiv \flow(\psi_S) =\unitperpmcurv$. 

A positivity analysis is now possible from considering equations 
\eqref{senwittenid}, \eqrefs{spatialdiraceq}{senwittensu2}, 
together with the fact that 
$\slDder{} = \slcovder{\Sigma}$
since $\Sigma_\perp$ is a maximal hypersurface. 
This yields the inequality 
\EQ
8\pi E(S;\sigma)
= 2 \int_{\Sigma_\perp} ( 
|\covDder{}\hyperpsi|^2 + \T{00}{} |\hyperpsi|^2 ) \d\Sigma
\geq 0
\doneEQ
whenever $\T{00}{} \geq 0$
(which is the weak energy condition \cite{HawkingEllis,Wald-book}
on the stress-energy tensor $\T{ab}{}$). 

Proposition~4:
Let $S$ be a convex twist-free spacelike 2-surface 
with positive Gaussian curvature, 
spanned by an adapted maximal spacelike hypersurface $\Sigma_\perp$ 
in spacetime $(M,g)$. 
If the spatial Dirac boundary value problem 
\eqref{spatialdiraceq}, \eqrefs{meancurvbc3}{meancurvbc2} 
possesses a smooth $SU(2)$ spinor solution $\hyperpsi$, 
then the mean-curvature mass \eqref{mcurvmass} is non-negative
provided the weak energy condition holds on $(M,g)$. 

Compared with Proposition~2, 
here a weaker energy condition is sufficient for positivity,
since the Dirac current \eqref{su2current} for $SU(2)$ spinors 
is hypersurface orthogonal,
but there is a stronger restriction on the 2-surface $S$.

\section{ Horizons and chiral boundary conditions }

In the positivity analysis, 
if a hypersurface $\Sigma$ contains an apparent horizon inside $S$
then the horizon 2-surface should be treated as an inner boundary 
$(\partial\Sigma)_\hor$,
which is characterized by the vanishing of its scalar mean curvature,
$\absmcurv_\hor=0$. 
On such hypersurfaces, 
the boundary value problem \sysref{witteneq}{meancurvbc2} 
needs to be supplemented by imposing the standard 
horizon boundary condition \cite{GHHP}
$\gmatr{}{1}\gmatr{}{0}\psi|_\hor =\psi|_\hor$
on the Dirac spinor in Witten's equation, 
in terms of an orthonormal frame adapted to $(\partial\Sigma)_\hor$. 

This boundary condition arises from the introduction of
chiral projection operators
\EQ
\proj{}{\pm} = \frac{1}{2}( \unitop \pm \gmatr{}{1} \gmatr{}{0} )
\doneEQ
characterized by the properties
$[\proj{}{\pm},\gmatr{}{\indS{a}}]=0$, 
$\proj{}{\pm}\gmatr{}{1}= \gmatr{}{1}\proj{}{\mp}$, 
$\proj{}{\pm}\gmatr{}{0}= \gmatr{}{0}\proj{}{\mp}$, 
in addition to 
$\proj{2}{\pm}= \proj{}{\pm}$, 
$\proj{}{+}\proj{}{-}=0$,
and $\proj{\ad}{\pm}= \proj{}{\pm}$. 
Relative to the decomposition $\psi=\psi_+ +\psi_-$
given by $\psi_\pm\equiv \proj{}{\pm}\psi$,
the surface integral terms in the Sen-Witten identity look like
\EQ
\int_S
2\psi_+^\ad \bslethD{}\psi_- +\cc +\absmcurv( |\psi_+|^2 + |\psi_-|^2 ) \d S
\doneEQ
after integration by parts with respect to $\bslethD{}$, 
discarding a total divergence on $S$ (by Stokes' theorem). 
Hence at a horizon $S=(\partial\Sigma)_\hor$,
the flux terms vanish under the boundary condition $\psi_-|_\hor=0$,
while the mean curvature terms are zero due to $\absmcurv_\hor=0$,
as first shown in \Ref{GHHP}. 
Consequently, the horizon boundary condition implies that 
the inner boundary makes no contribution to the quasilocal mass
so that the positivity analysis leading to propositions~2 and~4 
goes through as before.

\section{ Some concluding remarks on quasilocal spinorial mass }

It is natural to explore a spinor derivation and positivity analysis of
mean-curvature mass based on a chiral boundary condition
in place of the nonlinear boundary conditions investigated so far. 
Let $\hat\proj{}{\pm}$ denote the chiral projection operators adapted
to the mean curvature frame of a 2-surface $S$
and consider
\EQ\label{chiralbc}
\psi_-|_S = \phi_-|_S
\doneEQ
where, 
in a Euclidean embedding of $(S,\sigma)$
into a spacelike hyperplane in Minkowski space,
$\phi$ is a parallel Dirac spinor 
whose Dirac current is aligned with the timelike mean curvature vector of $S$,
\EQ\label{align}
\flow(\phi)= (\unitperpmcurv)_{\flt} .
\doneEQ
Note $[\hat\proj{}{\pm},\slSder{}]=0$ shows that $\phi_\pm$ obeys 
the boundary Witten equation
\EQ\label{witteneqSchiral}
\hat\proj{}{\mp} (\slcovder{S}\phi )_{\Mink}
= \gmatr{}{1}( \bslSder{}\phi_\mp +\frac{1}{2}\absmcurv_{\flt}\phi_\pm ) 
=0 .
\doneEQ
Also note, as a consequence of the alignment \eqref{align}, that
\EQ\label{norms}
|\phi_+|= |\phi_-| =\frac{1}{\sqrt 2}
\doneEQ
since 
$0=\flow^1(\phi)= (\unitmcurv)_\flt \cdot\flow(\phi)
= -\phi^\ad\gmatr{}{0}\gmatr{}{1}\phi
= \phi^\ad(\hat\proj{}{+}-\hat\proj{}{-})\phi
= |\phi_+|^2- |\phi_-|^2$
while 
$1= \flow^0(\phi)_\flt= |\phi|^2 
= |\phi_+|^2 + |\phi_-|^2$. 

Witten's equation \eqref{witteneq} 
with this chiral boundary condition \eqref{chiralbc}
is an elliptic boundary value problem for the Dirac spinor 
\EQ\label{chiralspinor}
\chi =\psi-\phi ,\quad
\chi_-|_S =0 ,
\doneEQ
satisfying the equation 
\EQ\label{chiraleq}
\slcovder{\Sigma}\chi = -\slcovder{\Sigma}\phi_- = -\Gamma\phi
\doneEQ
where $\Gamma$ is a certain gamma matrix operator
that vanishes in Minkowski space, $(\Gamma)_\Mink=0$. 
Existence and uniqueness of solutions $\chi$ is rigorously established
by the general results stated in \Ref{BartnikChrusciel}
for boundary value problems of this kind. 
It is important to note that, in particular, 
there are no zero modes for
the boundary value problem \sysref{chiralspinor}{chiraleq},
since if $\chi_0$ satisfies $\slcovder{\Sigma}\chi_0=0$
on a hypersurface $\Sigma$
with $\chi_-|_S =0$ at a boundary $S=\partial\Sigma$,
then the Sen-Witten identity evaluated for $\chi_0$
yields
$\int_\Sigma |\covder{\Sigma}\chi_0|^2 \d\Sigma = 
\int_\Sigma \flow(\chi_0)\cdot\T{0}{} \d\Sigma \leq 0$
provided the dominant energy condition holds on $\Sigma$.
This inequality implies $\covder{\Sigma}\chi_0=0$
so then $\chi_0|_\Sigma$ must be a spatially parallel spinor
which vanishes due to the chiral boundary condition. 

To proceed, let $\psi$ be the unique solution of 
the boundary value problem \sysref{chiralspinor}{chiraleq},
and consider the surface integral terms 
in the Sen-Witten identity \eqref{senwittenid}. 
Use of the boundary Witten equation \eqref{witteneqSchiral} for $\phi_-$
gives
$(\psi_+^\ad \slSder{}\psi_-)|_S
= (\psi_+^\ad \slSder{}\phi_-)|_S
= -\frac{1}{2} \absmcurv_\flt (\psi_+^\ad \phi_+)|_S$
for the flux terms. 
The complete Sen-Witten identity then yields the expression
\EQs
&& 8\pi\tilde E(S;\sigma,\psi) \equiv
\nonumber\\
&& \int_S( 
(\psi_+^\ad \phi_+ +\cc) \absmcurv_\flt
- (|\psi_+|^2 + |\phi_+|^2)\absmcurv
+(\psi_+^\ad\gmatr{}{0}\gmatr{}{\indS{a}}\phi_- +\cc)
\twistvec{}{\indS{a}} )\d S
\geq 0
\label{chiralid}
\doneEQs
assuming the dominant energy condition holds on $\Sigma$. 
The surface integral \eqref{chiralid} defines 
a purely spinorial quasilocal mass $\tilde E(S;\sigma,\psi)$
which can be naturally viewed as a
chiral variant of the mean-curvature mass of $S$.
Interestingly, this variant expression is rigorously positive
and should have some good properties as a quasilocal mass.
In any case, its positivity may perhaps provide a link
to the known positivity \cite{LiuYau} of mean-curvature mass. 

As it stands, however, this positivity property \eqref{chiralid}
gives only a lower bound
on the mean-curvature mass
\EQ
4\pi E(S;\sigma) = \frac{1}{2} \int_S ( \absmcurv_\flt -\absmcurv ) \d S
\geq 
-\int_S |\psi_+|^2 (\absmcurv_\flt -\absmcurv) +|\psi_+||\twist| \d S
\doneEQ
which follows from the elementary estimates
$\psi_+^\ad \phi_+ + \cc \leq |\psi_+|^2 + |\phi_+|^2$
and 
$\psi_+^\ad \gmatr{}{0}\gmatr{}{\indS{a}}\phi_- \twistvec{}{\indS{a}}
\leq |\psi_+| |\phi_-| |\twist|$,
combined with the normalizations \eqref{norms}. 
A slightly more satisfactory result can be obtained
if $S$ is restricted to be twist-free, 
in which case the twist term will drop out of the inequality \eqref{chiralid}.

Lemma~5:
Suppose $S$ is twist-free in spacetime
and let $\psi_+|_S$ be the boundary value of the solution 
of Witten's equation \eqref{witteneq} 
with the mean curvature chiral boundary condition \eqref{chiralbc},
under a Euclidean embedding of $S$. 
Then the mean curvature of $S$ obeys the inequality
\EQ
\int_S( (\smallfrac{1}{2}+|\psi_+|^2) (\absmcurv_\flt -\absmcurv) \d S
= 8\pi\tilde E(S;\sigma,\psi) 
\geq 0 .
\doneEQ

This inequality would directly imply positivity of the mean-curvature mass
$E(S;\sigma)$ for twist-free 2-surfaces $S$
if the boundary value $\psi_+|_S$ has constant norm on $S$.
A similar conclusion would follow from 
the more general inequality \eqref{chiralid}
under stronger conditions on $\psi_+|_S$
(namely, 
if $\psi_+|_S$ is equal to $\phi_+|_S$
so then 
$(\psi_+\gmatr{}{0}\gmatr{}{\indS{a}}\phi_-)|_S 
=\flow^{\indS{a}}(\phi_-)$
and 
$(\psi_+^\ad \phi_+)|_S = |\phi_+|^2 = |\phi_-|^2$,
implying the twist terms would vanish due to 
$(\psi_+\gmatr{}{0}\gmatr{}{\indS{a}}\phi_- +\cc)|_S \twistvec{}{\indS{a}} 
= \flow_\parallel(\phi)\cdot\twist=0$
while the mean curvature terms would simplify to 
$(|\psi_+|^2 + |\phi_+|^2)|_S (\absmcurv_\flt-\absmcurv)
= \absmcurv_\flt-\absmcurv$.) 
Clearly, such lines of argument 
will require a detailed analytical investigation of 
the mean curvature chiral boundary value problem for Witten's equation.

\begin{acknowledgments}
The author is supported by an NSERC grant. 
Roh Tung is thanked for stimulating conversations 
at an early stage of this research. 
\end{acknowledgments}

\appendix*

\section{ 2-spinor formulation }

Let $S$ be a spacelike 2-surface for which 
the mean curvature vector $\mcurv$ and its dual $\perpmcurv$ 
comprise an orthonormal frame 
$\e{0}{}=\unitperpmcurv,\e{1}{}=\unitmcurv$
in the normal space $\TperpS$. 
Fix a spin basis $\ospin{}{A},\ispin{}{A}$ 
\cite{Penrose-book}
aligned with mean curvature null frame 
\EQ\label{nullmcurvframe}
\e{}{AA'} = 
\ospin{}{A}\ocspin{}{A'} \nullmcurv{+} 
+ \ispin{}{A}\icspin{}{A'} \nullmcurv{-}
+\ospin{}{A}\icspin{}{A'} \m +\ispin{}{A}\ocspin{}{A'} \cm
\doneEQ
where 
\EQ
\ispin{}{A}\icspin{}{A'} \e{AA'}{} 
=\frac{1}{\sqrt{2}}( \unitperpmcurv +\unitmcurv ) \equiv \nullmcurv{+} ,\quad
\ospin{}{A}\ocspin{}{A'} \e{AA'}{} 
=\frac{1}{\sqrt{2}}( \unitperpmcurv -\unitmcurv ) \equiv \nullmcurv{-}
\doneEQ
are outgoing and ingoing null normals to $S$,
and 
\EQ
-\ispin{}{A}\ocspin{}{A'} \e{AA'}{} \equiv \m ,\quad
-\ospin{}{A}\icspin{}{A'} \e{AA'}{} \equiv \cm
\doneEQ
are linearly independent and tangential to $S$. 
Here the standard normalization $\ospin{A}{}\ispin{}{A}=1$ is employed. 
The extrinsic null mean curvature of $S$ is given by 
\EQ
\trk_\perp \equiv \trk(\nullmcurv{+}) = -\trk(\nullmcurv{-}) 
= \frac{1}{\sqrt{2}}\absmcurv
\doneEQ
from property \eqref{liedermcurv} of the mean curvature frame. 
Then, in the usual Newman-Penrose notation \cite{NP,Penrose-book}, 
\EQ
\rho \equiv \ospin{}{A} \covder{\cm} \ospin{A}{} =\trk_\perp ,\quad
\mu \equiv \ispin{}{A} \covder{\m} \ispin{A}{} = \trk_\perp 
\doneEQ
are the convergence/divergence of the outgoing/ingoing null normals at $S$
while
\EQ
\beta \equiv \ispin{}{A} \covder{\m} \ospin{A}{} +\cc 
= -\frac{1}{2} \twistvec{}{m} ,\quad
\beta' \equiv -\ospin{}{A} \covder{\cm} \ispin{A}{} +\cc 
= \frac{1}{2} \twistvec{}{\cm} 
\doneEQ
are the rotation (boost in the normal space) of the null normals
under infinitesimal displacement along the tangential directions of $S$. 
Note the equality $\rho=\mu$ 
distinguishes the geometrically preferred frame \eqref{nullmcurvframe}. 

A Dirac spinor consists of a pair of 2-spinors 
$(\spin{}{A},\varspin{A'}{})$
and reduces to a Majoranna spinor 
via the condition
$\varspin{A'}{}=\cspin{A'}{}$
which produces a complex-conjugate 2-spinor pair
$(\spin{}{A},\cspin{A'}{})$. 
Correspondingly,
the Dirac current is a timelike vector 
$\flowvec{AA'}{}(\spin{}{},\varspin{}{})
= \spin{}{A}\cspin{}{A'} + \varspin{}{A}\cvarspin{}{A'}$
for a Dirac spinor
and a null vector
$\flowvec{AA'}{}(\spin{}{}) = \spin{}{A}\cspin{}{A'}$
for a Majoranna spinor. 

In the mean curvature spin basis, 
write $\spin{}{A}=\psi_0\ospin{}{A}+ \psi_1\ispin{}{A}$. 
The Witten-Nester 2-form for a 2-spinor is given by
$\i \cspin{A'}{} \covder{}\spin{A}{} \wedge \e{}{AA'}$, 
and integration over $S$ 
then yields the surface integral terms in the Sen-Witten 2-spinor identity
\EQ
\int_S \i \cspin{A'}{} \covder{}\spin{A}{} \wedge \e{}{AA'}
= 
\int_S (
\trk_\perp ( |\psi_0|^2 + |\psi_1|^2 ) 
+\bar\psi_1 \eth\psi_0 +\cc ) \d S
\doneEQ
where $\eth$ is the standard edth operator \cite{edth,NP,Penrose-book}. 
Here 
\EQ
\frac{1}{\sqrt{2}} ( |\psi_0|^2 + |\psi_1|^2 ) 
= -\flow(\spin{}{})\cdot\unitperpmcurv
\doneEQ
is the Dirac norm of the 2-spinor at $S$
while
\EQ
\bar\psi_1 \eth\psi_0 +\cc 
= \flux(\spin{}{};\twist)
\doneEQ
is the tangential flux of the 2-spinor at $S$. 
Now the boundary Witten equation associated with $S$ in $M$ reduces to 
\EQ
\ispin{}{A}\covder{m}\spin{A}{} 
= \eth\psi_0 +\trk_\perp \psi_1 =0 ,\quad
\ospin{}{A}\covder{\bar m}\spin{A}{} 
= -\bar\eth\psi_1 +\trk_\perp \psi_0 =0 
\doneEQ
which is a 1st order PDE system for $\psi_0,\psi_1$. 
For the derivation of mean-curvature mass in theorem~1, however, 
note it is an embedded boundary Witten equation in Minkowski space 
that is used.

\def\v#1{{\bf #1}}

\end{document}